\documentclass[prc,twocolumn,amsmath,letterpaper]{revtex4}
\usepackage{graphicx}

\begin{document}

\title{Determining the Energy Barrier for Decay out of Superdeformed Bands} 

  \author{B. R. Barrett$^1$} 
  \author{J. B\"urki$^2$} 
  \author{D. M. Cardamone$^3$}
  \author{C. A. Stafford$^1$} 
  \author{D. L. Stein$^4$} 
  \affiliation{$^1$Physics Department, University of Arizona, Tucson, AZ
    85721\\
    $^2$Department of Physics and Astronomy, California State University,
    Sacramento, CA 95819\\
    $^3$Department of Physics and Astronomy, University of California, Irvine, CA 92697\\
  $^4$Department of Physics and Courant Institute of Mathematical Sciences, New York University, New York, NY 10003}

\begin{abstract}
An asymptotically exact quantum mechanical calculation of the matrix elements for tunneling through an asymmetric barrier
is combined with the two-state statistical model for decay out of superdeformed bands to determine
the energy barrier (as a function of spin) separating the superdeformed and normal-deformed wells for several nuclei in the 190 and
150 mass regions.  
The spin-dependence of the barrier leading to sudden decay out is shown to be
consistent with the decrease of a centrifugal barrier with decreasing angular
momentum.
Values of the barrier frequency in the two mass regions are predicted.
\end{abstract}

\maketitle

\section{Introduction}

Since their first experimental observation~\cite{twin86a}, superdeformed (SD)
nuclear states, with their strong ellipsoidal deformation and special set of
shell closures, have offered a tantalizing and unique window into subatomic
physics. Their rapid decay-out, in particular, has been the subject of great
interest~(e.g., Refs.~[2--15]\nocite{vigezzi90a,vigezzi90b,twin90a,Shimizu93,aberg99a,SB,wilson03a,dzyublik03a,CSBPRL,andreoiu03a,wilson04a,wilson05a,caurier05a,CBS08}). In the standard theoretical approach~\cite{vigezzi90a,vigezzi90b}, this process is modeled by a
two-well potential function of deformation: Here, the nucleus is a single
quantum mechanical particle, which tunnels between the two wells, and can
escape the system via electromagnetically induced decay from either. Because the
barrier between the SD and normal-deformed (ND) wells is a direct consequence
of nucleon--nucleon interactions, an understanding of its shape for various
nuclei and angular momenta would be of considerable importance to the study of
nuclear structure. Thus, a common objective of theoretical studies is to
bridge the gap between measured experimental data, such as lifetimes and
nuclear spins, and the shape of this barrier.
In this Letter, we show that a 
rapid
decrease in barrier height with decreasing nuclear spin
explains the SD decay mechanism.

\begin{figure}
  \centerline{\includegraphics[keepaspectratio=true,width=\columnwidth]{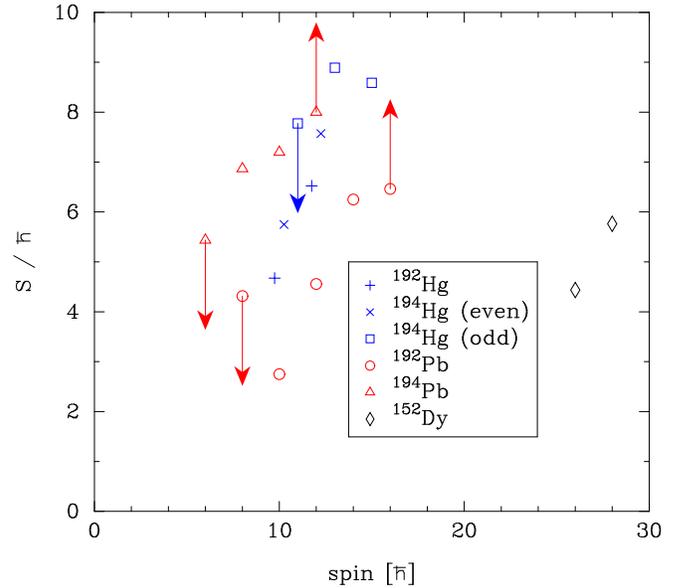}}
  \caption{
Calculated tunneling action $S=3.6 \,W/\omega$ versus spin for several SD
decays in the 190 and 150 mass regions. $W$ is the height of the energy
barrier, and $-\omega$ its curvature.
Arrows indicate cases for which only an upper or
lower bound on $F_N$ has been measured.
}
  \label{fig:action}
\end{figure}

It was previously shown \cite{CSBPRL}
that an elegant,
two-\emph{state} model~\cite{SB} of SD decay-out is sufficient to give an
excellent picture of the system's time-evolution. One of the principal
advantages of such a straightforward technique was the extraction from
experiment of such important quantities as the tunneling matrix element $V$
and the spreading width for tunneling through the barrier
$\Gamma^\downarrow$. The purpose of the present Letter is to move beyond
these phenomenological quantities, and extract the height of the barrier
itself as a function of nucleus and nuclear spin, as shown in
Fig.~\ref{fig:action}. 
Previous approaches \cite{kruecken96a} used a semiclassical treatment
that did not allow for an accurate computation of the tunneling rate
pre\-fac\-tor.  As we shall see, this can introduce a potential error of several
orders of magnitude in the estimation of the tunneling width.  In this
paper, we go beyond previous treatments by computing the tunneling rate in a
systematic and controlled fashion using a functional integral approach \cite{HTB90}.

\section{Path-integral approach to tunneling}

Absent additional information about the nuclear shell potential, the smoothest potential describing both the tunnel barrier 
and the SD well (which lies far above the ND yrast line at decay out) is a cubic polynomial:  
\begin{equation}
\label{eq:potential}
U(x) = {\cal M}\omega^2 x^2/2 - \lambda x^3,
\end{equation}
where $x$ is a coordinate describing the quadrupole deformation of the nucleus ($x=0$ corresponds to the bottom of the SD well, while the
ND well occurs for $x>x_B={\cal M}\omega^2/3\lambda$), $\omega$ is the oscillator frequency of the SD well,
and ${\cal M}$ is the inertia of the quadrupole vibrational mode.  
With a suitable choice of the parameters ${\cal M}\omega^2$ and $\lambda$, $U(x)$ provides a {\em maximum entropy} (least-biased) fit
to the unknown nuclear potential barrier.
Note that for this simple potential, the barrier frequency
$\omega_B\equiv \sqrt{|U''(x_B)|/{\cal M}}=\omega$.

The use of Euclidean complex-time path integrals over the tunneling
coordinate provides a systematic approach for determining quantum
tunneling rates at arbitrary temperature and dissipation~\cite{HTB90}, as
an asymptotic expansion in $\hbar$. This method allows calculation not only
of the leading-order exponential dependence of the tunneling rate on
potential parameters, but also the more computationally difficult
subdominant asymptotics (i.e., the prefactor term). For 
tunneling through the potential barrier~(\ref{eq:potential}) at zero
temperature and friction~[18--21]\nocite{CL83,FRH86a,FRH86b,S05}, the mean-square
tunneling matrix element out of the SD yrast state is found to be:
\begin{equation}
\langle V^2 \rangle = \hbar \omega D_N \left(\frac{54}{\pi^3}\frac{W}{\hbar\omega}\right)^{1/2} 
\!\!\! \exp\left(-\frac{36}{5}\frac{W}{\hbar\omega}\right),
\label{eq:gamma_dan}
\end{equation}
where $W\equiv U(x_B)={\cal M}^3 \omega^6/54\lambda^2$ is the barrier height
(as measured from the bottom of the SD well), and $D_N$ is the mean level
spacing in the ND band at the energy of the decaying SD state. 
The action $S$ to tunnel out of the SD state through the barrier is proportional
to the argument of the exponential function in Eq.~(\ref{eq:gamma_dan}):
\begin{equation}
-\frac{2S}{\hbar}=-\frac{36}{5}\frac{W}{\hbar\omega},
\end{equation}
where the factor of two is due to the power of $V$.

To make contact with experiment, the tunneling matrix element
may be estimated using the 
two-state model of SD decay \cite{SB,CSBPRL,CBS08}, which assumes the
decay-out process is
dominated by coupling of each SD state with its nearest-lying energy level in the ND band.  
The branching ratios $F_N$ and $F_S=1-F_N$ for decay out and intraband decay,
respectively, are determined by three rates \cite{CSBPRL}:
\begin{equation}
\label{eq:FN}
F_N  =  \frac{\Gamma_N\Gamma^\downarrow/\left(\Gamma_N+\Gamma^\downarrow\right)}
{\Gamma_S+\Gamma_N\Gamma^\downarrow/\left(\Gamma_N+\Gamma^\downarrow\right)},
\end{equation}
where $\Gamma_S/\hbar$ and $\Gamma_N/\hbar$ are the electromagnetic decay
rates of the SD and ND states, respectively, and
\mbox{$\Gamma^\downarrow/\hbar = \frac{2\bar{\Gamma} V^2}{\hbar\left(\Delta^2 +
  \bar{\Gamma}^2\right)}$}
is the nucleus' net tunneling rate through the barrier, with
 $\bar{\Gamma}=\frac{1}{2}(\Gamma_S+\Gamma_N)$, and 
$\Delta$ the energy difference between the 
SD and ND states~\cite{deltanote}.
Given the experimentally determined branching ratios and the electromagnetic widths, $\Gamma^\downarrow$ is known~\cite{CBS08}:
\begin{equation}
\label{eq:gammadown}
\Gamma^\downarrow=
\Gamma_S\bigg/\left(\frac{F_S}{F_N}-\frac{\Gamma_S}{\Gamma_N}\right).
\end{equation}
The tunneling matrix element $V$ may then be determined statistically \cite{CSBPRL,CBS08}, 
assuming the SD and ND levels are uncorrelated, and that the ND levels
obey the Wigner surmise.
The mean-square tunneling matrix element is found to be \cite{CBS08}
\begin{equation}
\langle V^2 \rangle  = D_N^2 \Gamma^\downarrow/6\pi \bar{\Gamma},
\label{eq:Vms}
\end{equation}
where a numerically negligible correction, whose relative size is ${\cal O}(\bar{\Gamma}/D_N)^2$, has been omitted. 

Eqs.\ (\ref{eq:gamma_dan}) and (\ref{eq:Vms}) may be combined to yield an expression for the tunneling width
in terms of the properties of the nuclear potential barrier:
\begin{equation}
\Gamma^\downarrow = \frac{18 \, \hbar \omega \, \bar{\Gamma}}{D_N}
\left(\frac{6W}{\pi\hbar \omega}\right)^{1/2} \!\!\! \exp\left(-\frac{36}{5}\frac{W}{\hbar \omega}\right).
\label{eq:barrier}
\end{equation}
Note that this result for the net tunneling width, which includes tunneling and electromagnetic decay on an equal footing,
differs by a factor of $3\bar{\Gamma}/D_N$ from the bare tunneling
width into an infinitely broadened, fully continuum ND spectrum.
From the values of $\Gamma_S$, $\Gamma_N$, and $D_N$ listed in Table \ref{bigtable}, one sees that
usage of such a bare ND-continuum result \cite{Shimizu93,khoo93a}
could result in an error of several orders of magnitude.

The energy barriers (in units of the barrier frequency) obtained by solving Eq.\ (\ref{eq:barrier}) 
for all SD decays for which the four parameters, $F_N$, $\Gamma_S$, $\Gamma_N$, and $D_N$, are known
are listed in Table \ref{bigtable} (see also Fig.\ \ref{fig:action}).
Also listed is the tunneling action $S=3.6 \, W/\omega$, 
which is a characteristic measure of the opaqueness of the barrier~\cite{Shimizu93,khoo93a}.
Note that $W/\hbar \omega$ depends only weakly (logarithmically) on the barrier frequency $\omega$. 
In the literature, the value \mbox{$\hbar\omega=0.6\mbox{MeV}$} has been used \cite{vigezzi90b,khoo93a}, but
we shall determine $\omega$ self-consistently in Section~\ref{sec:centrifugal}.

\begin{table*}
\caption{Barrier height $W$ and tunneling action $S$
  for all SD decays for which
  sufficient data (branching ratios, $\Gamma_S$, $\Gamma_N$, and $D_N$) are
  known. The rightmost column gives the sources of the experimental
  inputs as well as the sources of
  the estimates of $\Gamma_N$ and $D_N$.  The values of $\Gamma^\downarrow$ were calculated 
  using Eq.\ (\ref{eq:gammadown}), as discussed in Ref.~\cite{CBS08}.
  The barrier frequency $\omega$ was determined self-consistently in the 190 and 150 mass regions, 
  respectively, assuming the angular momentum dependence of the barrier height can be fit to that of a centrifugal barrier (see text).}
\label{bigtable}
\begin{tabular}{|c||r@{.}lr@{.}lr@{.}lr@{.}lr@{.}l|r@{.}lr@{.}lr@{.}l|c|}
\hline
nucleus(I) & \multicolumn{2}{c}{$F_N$} & \multicolumn{2}{c}{$\Gamma_S$} &
\multicolumn{2}{c}{$\Gamma_N$} & \multicolumn{2}{c}{$D_N$} &
\multicolumn{2}{c}{$\Gamma^\downarrow$} 
& \multicolumn{2}{c}{$\hbar \omega$}  
& \multicolumn{2}{c}{$W/\hbar \omega$}  
& \multicolumn{2}{c}{$S/\hbar$} & 
Refs.\\
& \multicolumn{2}{c}{} 
& \multicolumn{2}{c}{(meV)} & \multicolumn{2}{c}{(meV)}
& \multicolumn{2}{c}{(eV)} & \multicolumn{2}{c}{(meV)} & \multicolumn{2}{c}{(MeV)} & \multicolumn{2}{c}{} & \multicolumn{2}{c}{} & \\
\hline
\hline
${}^{192}$Hg(12) & 0&26  & 0&128 & 0&613 & 135& & 0&049 & 0&24 & 1&8 & 6&5 & \cite{lauritsen00a,wilson05a}\\ 
${}^{192}$Hg(10) & 0&92  & 0&050 & 0&733 & 89& & 0&37 & 0&24 & 1&3 & 4&7 & \cite{lauritsen00a,wilson05a}\\ 
\hline
${}^{192}$Pb(16) & $<$0&01 & 0&487 & 0&192 & 1,362& & $<$0&0050 & 0&24 & $>$1&8 & $>$6&5 & \cite{wilson03a,wilson04a}\\ 
${}^{192}$Pb(14) & 0&02 & 0&266 & 0&201 & 1,258& & 0&0056 & 0&24 & 1&7 & 6&2 & \cite{wilson03a,wilson04a}\\ 
${}^{192}$Pb(12) & 0&34 & 0&132 & 0&200 & 1,272& & 0&10 & 0&24 & 1&3 & 4&6 & \cite{wilson03a,wilson04a}\\ 
${}^{192}$Pb(10) & 0&88 & 0&048 & 0&188 & 1,410& & 1&9
& 0&24 & 0&76 & 2&7 & \cite{wilson03a,wilson04a}\\ 
${}^{192}$Pb(8) & $>$0&75 & 0&016 & 0&169 & 1,681& & $>$0&067 & 0&24 & $<$1&2 & $<$4&3 & \cite{wilson03a,wilson04a}\\ 
\hline
${}^{194}$Hg(12) & 0&42 & 0&097 & 4&8 & 16&3 & 0&071 & 0&24 & 2&3 & 8&4 &\cite{khoo96a,kuehn97a,moore97a,kruecken01a}\\ 
${}^{194}$Hg(10) & $>$0&91 & 0&039 & 4&1 & 26&2 & $>$0&44 & 0&24 & $<$2&0 & $<$7&1 & \cite{khoo96a,kuehn97a,moore97a,kruecken01a}\\ 
\hline
${}^{194}$Hg(12) & 0&40 & 0&108 & 21& & 344& & 0&072 & 0&24 & 2&1 & 7&6 & \cite{lauritsen02a}\\ 
${}^{194}$Hg(10) & 0&97 & 0&046 & 20& & 493& & 1&6 & 0&24 & 1&6 & 5&8 & \cite{lauritsen02a}\\ 
\hline
${}^{194}$Hg(12) & 0&40 & 0&086 & 1&345 & 19& & 0&060 & 0&24 & 2&2 & 7&8 & \cite{moore97a,wilson05a}\\ 
${}^{194}$Hg(10) & $\ge$0&95 & 0&033 & 1&487 & 14& & $\ge$1&1 & 0&24 & $\le$1&8 & $\le$6&5 & \cite{moore97a,wilson05a}\\ 
\hline
${}^{194}$Hg(15) & 0&10 & 0&230 & 4&0 & 26&5 & 0&026 & 0&24 & 2&4 & 8&6 & \cite{moore97a,kruecken01a}\\ 
${}^{194}$Hg(13) & 0&16 & 0&110 & 4&5 & 19&9 & 0&021 & 0&24 & 2&5 & 8&9 & \cite{moore97a,kruecken01a}\\ 
${}^{194}$Hg(11) & $>$0&93 & 0&048 & 6&4 & 7&2 & $>$0&71 & 0&24 & $<$2&2 & $<$7&8 & \cite{moore97a,kruecken01a}\\ 
\hline
${}^{194}$Pb(10) & 0&10 & 0&045 & 0&08 & 21,700& & 0&0053 & 0&24 & 1&1 & 4&1 & \cite{willsau93a,lopezmartens96a,hauschild97a,kruecken01a}\\ 
${}^{194}$Pb(8) & 0&38 & 0&014 & 0&50 & 2,200& & 0&0087 & 0&24 & 1&6 & 5&8 & \cite{willsau93a,lopezmartens96a,hauschild97a,kruecken01a}\\ 
${}^{194}$Pb(6) & $>$0&91 & 0&003 & 0&65 & 1,400& & $>$0&032 & 0&24 & $<$1&5 & $<$5&5 & \cite{willsau93a,lopezmartens96a,hauschild97a,kruecken01a}\\ 
\hline
${}^{194}$Pb(12) & $<$0&01 & 0&125 & 0&476 & 236& & $<$0&0013 & 0&24 & $>$2&2 & $>$8&0 & \cite{kruecken01a,wilson04a}\\ 
${}^{194}$Pb(10) & 0&10 & 0&045 & 0&470 & 244& & 0&0051 & 0&24 & 2&0 & 7&2 & \cite{kruecken01a,wilson04a}\\ 
${}^{194}$Pb(8) & 0&35 & 0&014 & 0&445 & 273& & 0&0077 & 0&24 & 1&9 & 6&9 & \cite{kruecken01a,wilson04a}\\ 
${}^{194}$Pb(6) & $>$0&96 & 0&003 & 0&405 & 333& & $>$0&088 & 0&24 & $<$1&5 & $<$5&4 & \cite{kruecken01a,wilson04a}\\ 
\hline\hline
${}^{152}$Dy(28) & 0&40 & 10&0 & 17& & 220& & 11& & 0&56 & 1&6 & 5&8 & \cite{lauritsen02a}\\ 
${}^{152}$Dy(26) & 0&81 & 7&0 & 17& & 194& & 140&
& 0&56 & 1&2 & 4&4 & \cite{lauritsen02a}\\ 
\hline
\end{tabular} 
\end{table*}

We note that, of the four parameters in Eq.~(\ref{eq:barrier}), only $F_S$ and
$\Gamma_S$ are directly measured experimentally; typically, these are known to within
a few percent. $\Gamma_N$ and $D_N$ must be calculated
theoretically, with models fit to experimental data. The
uncertainties in $\Gamma_N$ and $D_N$ could thus be appreciable, perhaps as
large as a factor of two or more. A better determination of these quantities
is a worthy goal for future studies of SD nuclei, but goes well beyond the
scope of the present Letter.

For almost all decay-out sequences, we find that the barrier height decreases with decreasing angular momentum.  One exception is the odd-spin
$^{194}$Hg sequence, for which the two highest-spin calculated barriers are so close that statistical fluctuations about the mean-square matrix
elements of Eqs.\ (\ref{eq:gamma_dan}) and (\ref{eq:Vms}) are sufficient to reverse the trend.  This could occur, for example, due to an
accidental near-degeneracy of the SD and ND states in $^{194}$Hg(15), which
would lead to a larger than expected branching ratio $F_N$. The other
exception is the first parameter set for
${}^{194}\mathrm{Pb}(10)$. The primary difference between the first and second
parameter sets for ${}^{194}\mathrm{Pb}$ is Ref.~\cite{wilson04a}'s revised
treatment of the pairing gap; it is thus seen that this
consideration may play an important role in understanding the spin dependence
of the SD decay-out barrier~\cite{awnote}.

\section{Centrifugal tunnel barrier}
\label{sec:centrifugal}

Finally, we address whether the decrease in the tunnel barrier with decreasing spin (cf.\ Fig.\ \ref{fig:action} and Table \ref{bigtable})
is consistent with the centrifugal barrier of a spinning nucleus.
If the superdeformed nucleus and the saddle configuration at the top of the energy barrier
are described as rigid rotors with moments of inertia ${\cal I}_S$ and ${\cal
  I}_B$, respectively, 
then the barrier height $W(I)$, as a function of the angular momentum quantum number $I$,
is simply 
the $I=0$ barrier $W(0)$, 
plus the rigid-rotor rotational increase in the energy of the barrier configuration, \emph{minus} the
rotational increase of the bottom of the SD well (from which $W(I)$ is
measured), i.e.,
\begin{equation}
W(I) = W(0) + \frac{\hbar^2 I (I+1)}{2} \left(\frac{1}{{\cal I}_B} - \frac{1}{{\cal I}_S}\right).
\end{equation}
Although the rigid-rotor model is 
a simplification,
nevertheless the
decrease in the barrier height between successive SD states can be rigorously 
expressed in terms of the kinetic moments of inertia:
\begin{equation}
W(I) - W(I-2) = \hbar^2 (2I-1) \left(\frac{1}{{\cal I}_B^{(1)}} - \frac{1}{{\cal I}_S^{(1)}}\right).
\label{eq:barriervsI}
\end{equation}
The kinetic moments of inertia ${\cal I}_S^{(1)}$ of several SD yrast states in the 150 and 190 mass regions have been measured.
For $^{152}$Dy, ${\cal I}_S^{(1)}=85 \hbar^2/\mbox{MeV}$ 
and the aspect ratio $\eta \equiv b/a=2.0$ \cite{lauritsen02a},
with $a$ and $b$ the smaller and larger radii of the nucleus, respectively.
For $^{192}$Hg, ${\cal I}_S^{(1)}=90 \hbar^2/\mbox{MeV}$ and 
the aspect ratio $\eta=1.65$ \cite{janssens91}.

The moment of inertia of the barrier configuration ${\cal I}_B^{(1)}$ is not measured, but must be determined theoretically.  This could be done by applying
the Strutinsky shell correction method to the cranking model \cite{strutinsky68a}.
However, to account for pairing, we employ a phenomenological two-fluid model~\cite{shapesandshells}
in which only the region outside the largest possible central
sphere contributes to the moment of inertia.
Within this two-fluid model, we find that the moment of inertia is \cite{Barrett09}
\begin{equation}
{\cal I}^{(1)} = m_n r_0^2 \left(\frac{A}{\eta}\right)^{5/3} \frac{\eta^3 + \eta- 2}{5},
\label{eq:2fluid}
\end{equation}
where the nucleus has been taken as a prolate ellipsoid of revolution with
aspect ratio $\eta$ and atomic mass number $A$,
$m_n$ is the mass of a nucleon, and $r_0 = 1.27\mbox{fm}$.  With these parameters, the measured kinetic moments of
inertia of $^{152}$Dy and $^{192}$Hg at decay-out are reproduced to within 1\
To leading order in the quadrupole deformation parameter $\varepsilon$ (see 
Ref.~\cite{shapesandshells}), Eq.~(\ref{eq:2fluid}) gives
\begin{equation}
{\cal I}^{(1)} 
\approx 
\frac{4\varepsilon}{5} 
A^{5/3} m_n r_0^2.
\end{equation}

For $^{152}$Dy, the barrier occurs at an aspect ratio of $\eta=1.7$ \cite{nolan88}, so that ${\cal I}_S^{(1)}/{\cal I}_B^{(1)}=1.3$
and $\Delta W=W(28)-W(26)=0.21\mbox{MeV}$.  Assuming a constant barrier frequency, and comparing to the results from Table \ref{bigtable}
($\Delta W/\hbar\omega=0.37$), implies a barrier frequency $\hbar \omega=0.56\mbox{MeV}$.

For $^{192}$Hg, the barrier is estimated to occur at an aspect ratio of $\eta\approx 1.4$ \cite{janssens91}, 
so that ${\cal I}_S^{(1)}/{\cal I}_B^{(1)}=1.5$
and $\Delta W=W(12)-W(10)=0.12\mbox{MeV}$.  Assuming a constant barrier frequency, and comparing to the results from Table \ref{bigtable}
($\Delta W/\hbar\omega=0.51$), implies a barrier frequency $\hbar
\omega=0.24\mbox{MeV}$. Because the logarithmic dependence of $W$ on $\omega$
almost completely cancels out in such a calculation, the differences $\Delta
W/\omega$ and $\Delta S$ are nearly independent of the choice of $\omega$.

\section{Conclusions}

In conclusion, we have determined
the barrier height $W$ and tunneling action $S$
for decay-out of a superdeformed band by combining 
an asymptotically exact quantum tunneling calculation with 
a two-state dynamical model.
The Table
presents our numerical results for all superdeformed decays for which
sufficient experimental data are known. We find that
the tunnel barrier decreases significantly
with decreasing spin during the decay-out process. 
The spin-dependence of the barrier is 
explained quantitatively in terms of the variation of
the centrifugal barrier within a two-fluid model of nuclear rotation,
which in turn allows us to self-consistently predict the tunnel
barrier's curvature.

The results presented in this Letter complete the chain of reasoning needed to connect the intriguing phenomenology
of the decay-out process in superdeformed nuclei with an understanding of the underlying nuclear structure.
Our results indicate that the rapidity and universality of the decay-out profiles
can be explained straightforwardly within our two-state dynamical model by the decrease 
of the centrifugal barrier between the super-deformed and normal-deformed energy wells with decreasing spin.

\section*{Acknowledgments}
The authors thank Teng Lek Khoo and Anna Wilson for useful discussions; 
TRIUMF, the GSI Helmholtzzentrum f\"ur Schwerionenforschung, and the Institute for Nuclear Theory at the University of Washington for their hospitality;
and the Department of Energy for partial support during the completion of this work.
BRB was supported in part by NSF grant PHY-0555396 and the Alexander von
Humboldt Stiftung.  DLS was supported in part by NSF grant PHY-06501077.

\bibliographystyle{elsarticle-num}
\bibliography{sd_barrier}

\begin{thebibliography}{10}
\bibitem{deltanote}
$\Delta$ should not be \emph{directly} associated with the experimentally
  observed energy splitting, since the SD and ND states are not eigenstates of
  the full Hamiltonian. Rather, an observed energy splitting may be assocaited
  with $\sqrt{\Delta^2+4V^2}$ (see Ref.~\cite{SB}).
\end{thebibliography}

\end{document}